\documentclass[11pt,review]{elsarticle}
\biboptions{numbers,sort&compress} % source: http://tex.stackexchange.com/questions/82293/ddg#82294
\journal{arXiv}
\usepackage{hyperref}
\usepackage[utf8]{inputenc}
\usepackage{graphicx}
\usepackage{booktabs} % for tables
\usepackage{xcolor}
\usepackage{amsmath, amssymb, amsthm} % for proof environment
\allowdisplaybreaks
\newcommand{\RX}{\mathbb R} % real numbers
\newcommand{\EQNREF}[1]{(Eq.~\ref{#1})}

\newtheorem{theorem}{Theorem}
\newcommand{\THEREF}[1]{Theorem~\ref{#1}}

\makeatletter
\let\@afterindenttrue\@afterindentfalse
\makeatother

\begin{document}

\begin{frontmatter}

\title{Social Cost of Carbon: What Do the Numbers Really Mean?}

\author[addr-iiasa-em]{Nikolay Khabarov\corref{tag-correspondingauthor}}
\cortext[tag-correspondingauthor]{Corresponding author}
\ead{khabarov@iiasa.ac.at} 

\author[addr-iiasa-ibf,addr-msu]{Alexey Smirnov}
\author[addr-iiasa-em,addr-oxford]{Michael Obersteiner}

\address[addr-iiasa-em]{Exploratory Modeling of Human-Natural Systems Research Group, 
Advancing Systems Analysis Program,
International Institute for Applied Systems Analysis (IIASA), 
Schlossplatz 1, Laxenburg, A-2361, Austria}

\address[addr-iiasa-ibf]{Integrated Biosphere Futures Research Group,  Biodiversity and Natural Resources Program,
International Institute for Applied Systems Analysis (IIASA), 
Schlossplatz 1, Laxenburg, A-2361, Austria}

\address[addr-msu]{Faculty of Computational Mathematics and Cybernetics,
Lomonosov Moscow State University, Moscow, 119991, Russia}

\address[addr-oxford]{Environmental Change Institute, Oxford University Centre for the Environment, South Parks Road, Oxford, OX1 3QY, UK}

\begin{abstract}
%\begin{linenumbers}
Social cost of carbon (SCC) is estimated by integrated assessment models (IAM) and is widely used by government agencies to value climate policy impacts.
While there is an ongoing debate about obtained numerical estimates and related uncertainties, little attention has been paid so far to the SCC calculation method itself.

This work attempts to fill the gap by providing theoretical background and economic interpretation of the SCC calculation approach implemented in the open-source IAM DICE (Dynamic Integrated model of Climate and the Economy).
Our analysis indicates that the present calculation method provides an approximation that might work pretty well in some cases, while in the other cases the estimated value substantially (by the factor of four) deviates from the ``true'' value.
	\newpage % temporary here in the draft: abstract overlapping the bottom heading
This deviation stems from the inability of the present calculation method to catch the linkages between two key IAM's components -- complex interconnected systems -- climate and economy, both influenced by emission abatement policies. 
Within the modeling framework of DICE, the presently estimated SCC valuates policy-uncontrolled emissions against economically unjustified consumption, which makes it irrelevant for application in climate-economic policies and, therefore, calls for a replacement by a more appropriate indicator.

	An apparent SCC alternative, which can be employed for policy formulation is the direct output of the DICE model -- the socially optimal marginal abatement cost (SMAC), which corresponds to technological possibilities at optimal level of carbon emissions abatement.
In policy making, because of the previously employed implicit approximation, great attention needs to be paid to the use of SCC estimates obtained earlier.

	\phantom{space-for-keywords} 

%\end{linenumbers}
\end{abstract}


\end{frontmatter}

\section*{One sentence summary} 

The concept of the Social Cost of Carbon (SCC) as manifested through its estimation methodology and widely used in policy making needs to be replaced, because it valuates policy-uncontrolled emissions against economically unjustified consumption, and which can substantially deviate from the ``true'' (intended) value of SCC in particular policy contexts.

\section{Introduction}

The concept of the Social Cost of Carbon (SCC) appeared in the early publications of Nordhaus~\cite{refNordhaus2019} and dates back to the first works on the Dynamic Integrated Climate-Economy (DICE) model~\cite{refDICEWebPage}.
SCC gained momentum for policy making in 2000's~\cite{refPearce2003} and since then was widely used by a large number of organizations e.g. Worldbank~\cite{refWorldbank2017}, US EPA~\cite{refEPA2010}, and UK DEFRA~\cite{refPearce2003}.
While according to more recent publications by Nordhaus~\cite{refNordhaus2019} the SCC did not play a decisive role in the evaluation of the US climate related policies, an earlier publication by Nordhaus~\cite{refNordhausPNAS2017} reported ``regulations with more than \$1 trillion of benefits have been written for the United States that use the SCC in their economic analysis''.
The SCC concept is well integrated in the current policy context and therefore plays an important role in assessments of climate related action.
The United States Government's Interagency Working Group on Social Cost of Carbon is using the SCC according to respective regulation~\cite{refEPA2010} relying for the purposes of the SCC estimation on the FUND\footnote{Climate Framework for Uncertainty, Negotiation and Distribution (FUND) Model.}~\cite{refFUND2013,refFUNDGitHub} and PAGE\footnote{Policy Analysis of the Greenhouse Effect (PAGE) Model.}~\cite{refPAGE2008,refPAGE09,refMimiPAGE,refPAGEGitHub} models along with the DICE model.
Recent publications hinge on the previously developed methodology:
there are calls for a modular modeling approach~\cite{refNatlAcad} and recommendations for a roadmap on improving numerical SCC estimates embedded in the policy context~\cite{refWagnerArnthoff}.

There are few definitions of SCC in the literature e.g. ``the social cost of carbon refers to the estimate of the monetary value of world-wide damage done by anthropogenic CO$_2$ emissions''~\cite{refPearce2003}, ``\phantom{`}`social cost of carbon' is defined as the monetary value of the damage done by emitting one more tonne of carbon at some point of time''~\cite{refPearce2003}, ``it is the change in the discounted value of economic welfare from an additional unit of CO$_2$-equivalent emissions''~\cite{refNordhaus2019}, ``it is the change in the discounted value of the utility of consumption per unit of additional emissions, denominated in terms of current consumption''~\cite{refNordhaus2014}, or ``SCC estimates the dollar value of reduced climate change damages associated with a one-metric-ton reduction in carbon dioxide (CO$_2$) emissions''~\cite{refPizer2014} to name a few.
Here, among the definitions, Pearce et al.~\cite{refPearce2003} are more explicit on the intended meaning of SCC by stating the {\it anthropogenic} nature of the emissions (as these are supposed to be subject of climate policies where SCC is employed), while other formulations are slack on this.

The DICE model can maximize social utility by finding an optimal level of carbon emissions abatement and---corresponding to that level---socially optimal marginal abatement cost (SMAC)\footnote{%
SMAC is the cost of one additional ton CO$_2$ reduction at the optimal abatement level in a particular year, which is a decision variable denoted in DICE GAMS source code as MUI(t). SMAC is denoted in the source code as MCABATE(t)  ``marginal cost of abatement'' and additionally as CPRICE(t) ``carbon price''. SMAC is calculated as $c_1(t) \cdot [ MIU(t) ]^{c_2}$, where $c_1(t)$ and $c_2$ are known model parameters.
}, which is the direct output of DICE.
As opposed to SMAC, SCC is an additional calculation on top of DICE outputs -- it is a ratio between the so called ``marginal'' values corresponding to model's emissions and consumption equations\footnote{The compact mathematical formulation of the DICE model is presented in Appendix A:  Simplified DICE formulation.}. As explained by Nordhaus, 
``the ratio calculates the economic impact of a unit of emissions in terms of t-period consumption as a num\'eraire''~\cite{refNordhaus2014}.

\begin{figure}[h]
        \centering
        \includegraphics[width=12 cm]{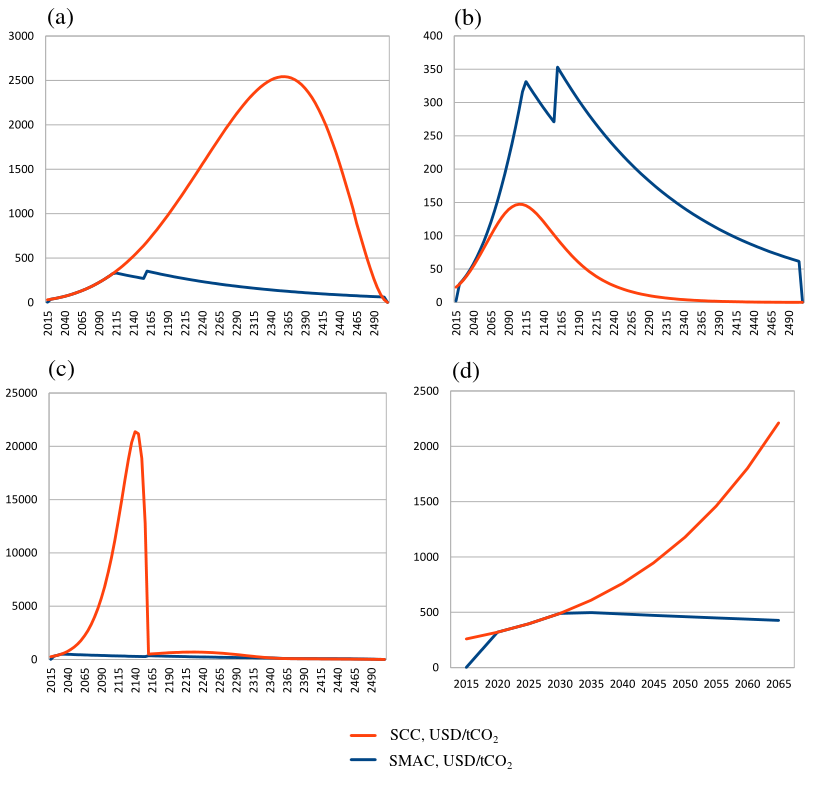}
	\caption{Social cost of carbon (SCC) and socially optimal marginal abatement cost of carbon (SMAC) as estimated by (a) unmodified DICE 2016 model, (b) DICE 2016 with the utility function replaced by utility not weighted by population size, (c) DICE 2016 model with added temperature constraint T {\textless} 2.4 \textsuperscript{o}C, (d) same as (c) yet zoomed into a shorter 50-year time period 2015-2065.}
    \label{fig1}
\end{figure}

A standard DICE 2016 model\footnote{Source: http://www.econ.yale.edu/\~{}nordhaus/homepage/homepage/DICE2016R-091916ap.gms (accessed on October 23, 2019).} run produces vividly different SCC and SMAC for the tail of the trajectory, see Figure~\ref{fig1}~(a).
The difference between SCC and SMAC, generally speaking, is not confined to the tail of the optimal trajectory, so the same model, with the exception that the utility function is replaced with the one that is not weighted by population size in different time periods~\cite{refScovronick2017}, produces visibly different SCC and SMAC also at the head of the optimal trajectory, see Figure~\ref{fig1}~(b).
The difference between SCC and SMAC can be positive or negative as evidenced by these figures, and the moment in time when this difference becomes noticeable can be rather close to the beginning of the modeling time interval as evidenced by results obtained from a modified version of the DICE 2016 model where only temperature constraint T {\textless} 2.4 \textsuperscript{o}C is added and the rest of the model is kept unchanged\footnote{The application of a direct temperature constraint is justified by the property of the damage function, which is unable to capture (or just translate to a monetary value) all potential damage stemming from increased GHG concentrations in the atmosphere.
Such constraint was applied and reported in DICE-2013 (Introduction and User Manual) and later also in DICE-2016 (%
%https://data.nber.org/reporter/2017number3/nordhaus.html
http://web.archive.org/web/20191205041047/ https://data.nber.org/reporter/2017number3/nordhaus.html%
).}, see Figure~\ref{fig1}~(c) and (d).

While SMAC is a direct result of the optimization, SCC is a result of an ad-hoc calculation, yet both correspond to the same optimal solution of the model.
Both SMAC and SCC are expressed in same units of US dollars per ton of CO$_2$ and represent cost of carbon associated respectively with abating or emitting one ton of CO$_2$ at an optimal level of abatemet in a particular time period. 
Intuitively, one may expect that at an optimal level of abatement these costs will be equal\footnote{%
	Otherwise, making an assumption that e.g. abatement is more expensive than emitting i.e. SMAC is greater than SCC, the last ton of CO$_2$ can be saved from abatement i.e. released to the atmosphere and the resulting cost will become SCC, which is less than the initial SMAC i.e. the new level of abatement would be better and hence the initial level is not optimal, which is a contradiction, because the optimal case is being considered. This contradiction makes the initially made assumption (SMAC $>$ SCC) invalid. Similar arguments lead to the conclusion that the case SMAC $<$ SCC is also impossible, hence, SMAC = SCC. Further in the analysis we do not appeal to this intuition -- it only serves the purpose of nurturing reader's interest.
}%
, which is not the case as shown in Figure~\ref{fig1}.
So, the challenging questions for the analysis that follows are: (a) why there is a numerical difference between SMAC and SCC and (b) what implications does this difference have for policymaking.

Unfortunately, the literature does not say anything clear on that and therefore cannot help answering these two questions, e.g.: ``With an optimized climate policy (abstracting away from complications due to tax or regulatory distortions or inconsistent treatment in different sectors), the {\it SCC will equal the carbon price}; this in turn is equal to the marginal cost of emissions reduction''~\cite{refNordhaus2014}, which complies with the intuition referred to earlier, but contradicts the fact of the difference between SCC and SMAC shown in Figure~\ref{fig1}; or, (in a cost--benefit analysis (CBA) model) ``the marginal social cost of carbon is the marginal damage done at the optimal level of abatement''~\cite{refPearce2003}, which refers to a ``damage'' and therefore does not inform on the relation between SCC and SMAC.
Despite the lack of clarity, policy makers are keen on employing SCC for various reasons. 
Therefore, it is of paramount importance to clarify the meaning of SCC as manifested by its estimation methodology.

\section{Definition and calculation}

%
%As these two marginal values are just the Lagrange multipliers~\cite{refSimonBlume1994} in DICE's constrained utility maximization problem\footnote{See \MARKIT{Appendix A}: Simplified DICE formulation and \MARKIT{Appendix B}: DICE and the Standard constrained optimization problem.}, they describe\footnote{See \MARKIT{Appendix C}: interpretation of marginal values in DICE.} by how much the optimal utility would change if one ton of CO$_2$ would be added to the emission balance equation and $x$ dollars would be added to the consumption equation and the new problem would be solved.
%These marginal values are expressed respectively in (a) units of utility per one ton of CO$_2$ and (b) units of utility per one dollar of additional consumption\footnote{Both values being Lagrange multipliers assume that the corresponding balance equation is perturbed (modified) by a ``sufficiently small'' quantity. It is assumed that one ton of CO$_2$ and $x$ dollars in this example satisfy that requirement.}.
%

In the literature, the DICE's direct output---socially optimal marginal abatement cost, SMAC,---is largely ``shadowed out'' by SCC, which is a combination of two indirect products of the model -- two so called ``marginal'' values that correspond to two specific model equations and stem from the computational method of finding a solution of the optimization problem. 
These two marginal values correspond to (a) emissions equation and (b) consumption equation in the mathematical formulation of the DICE optimization problem\footnote{For details see Appendix A: Simplified DICE formulation.} -- they are denoted in the DICE GAMS\footnote{https://www.gams.com} source code respectively as (a) $eeq.m(t)$ and (b) $cc.m(t)$, where $t$ indicates time period.
The equation for the SCC calculation as implemented in DICE is\footnote{The equation is adapted for clarity from the DICE source code by removing scaling and regularization factors.} 

\begin{equation}
	eeq.m(t) + x \cdot cc.m(t) = 0, \label{bkm:eq1}
\end{equation}
where marginal values $eeq.m(t)$ and $cc.m(t)$ are known and $x$ is the to-be-derived SCC value in the time period $t$. 

The GAMS documentation\footnote{URL: https://www.gams.com/33/docs/UG\_Glossary.html accessed on 2021-10-04.} explains the meaning of such marginal values and the notation used to refer to equations' marginal values in the GAMS system:

\begin{quote}
	``Marginal values (aka "dual values", "reduced costs", "shadow prices", or "multipliers") are stored in the ".m" [...] equation attribute. The GAMS sign convention is this: the marginal value represents the amount and direction of {\it change in the objective value given a unit increase in the binding constant} ([...] right-hand side [of an equation]).''
\end{quote}

So, the two marginal values, which are terms in the left-hand side of~\EQNREF{bkm:eq1}, have the following meaning: $eeq.m(t)$ -- is the increment of the objective value (i.e. optimal value of DICE's objective function -- social utility) corresponding to one unit (i.e. one ton CO$_2$) increase in right-hand side of the emissions equation and $cc.m(t)$ -- is the increment of the objective value corresponding to one unit (i.e. one dollar) increase in right-hand side of the consumption equation, this one dollar value is scaled to $x$ dollars increase in right-hand  side of the consumption equation giving $x \cdot cc.m(t)$ increment in objective value.

In summary, the left-hand side of~\EQNREF{bkm:eq1} represents the total increment of the objective value in a new problem as compared to the original DICE problem. 
The new optimization problem (further referred to as "perturbed problem") differs from original in two equations: (a) the emission equation is perturbed (modified) by adding one ton of CO$_2$ to its right-hand side, and (b) the consumption equation is perturbed by adding $x$ dollars to its right-hand side\footnote{
	For a detailed mathematical explanation of the link between the DICE model formulation, considered marginal values, and SCC equation \EQNREF{bkm:eq1}, please see the appendices -- Appendix A: Simplified DICE formulation,  Appendix B: DICE and the standard constrained optimization problem, and Appendix C: Interpretation of marginal values in DICE.
}.

Since this increment in the objective value of perturbed problem according to~\EQNREF{bkm:eq1} is equal to zero, there would be no change in objective value if the original DICE problem would be substituted by the perturbed problem i.e. the objective values in the original and the perturbed problems are equal.

Hence, the SCC equation~\EQNREF{bkm:eq1} means that addition of one ton of CO$_2$ to the right-hand side of the emissions balance equation and simultaneous addition of $x$ dollars to the right-hand side of the consumption equation would lead to a new optimization problem that has the same optimal value of social utility (objective function) as the original problem, i.e. one ton of added CO$_2$ emissions is being compensated by $x$ dollars of added consumption.
This allows one to call $x$ an ``exchange rate'' between additional emissions and additional consumption that keeps the ``status quo'' in terms of utility remaining constant.
The ``exchange rate'' can be seen as a monetary value compensating extra one ton of emissions to keep the societal ``status quo'', which justifies the name SCC.

\section{Interpretation}

Here we provide the interpretation of the perturbed problem, which is implicitly employed for the SCC calculation through the use of marginal values; it's derived from the original problem by modifying its emissions and consumption equations.
We start with a discussion on the meaning of the correction of the emissions equation by one ton of CO$_2$ in a particular year. 

The industry may {\it decide} for whatever reason to emit ``just a bit'' more\footnote{%
For the ease of storytelling, we consider adding emissions and refer to the compensating added consumption.
A similar consideration of reducing emissions and its compensating adjusted consumption is also valid.
} than planned (whether the plan is optimal or not), however, that would imply that the abated quantity\footnote{%
The related decision variable is denoted as the emission control rate, MIU(t) in the DICE 2016 GAMS source code.
} and/or the capital investment\footnote{%
The related decision variable is denoted as the gross savings rate as fraction of gross world product, S(t) in the DICE 2016 GAMS source code. Savings rate is a direct equivalent of capital investment in DICE as the unconsumed share of the economic product is ``saved'' by investing into capital.
} (both are the only {\it decision} variables in DICE\footnote{%
See Appendix A: Simplified DICE formulation.
}) should change so that the total production and associated emissions go up (or just emissions if only abatement level is reduced).
Therefore, in case of change in human-controlled emissions (in DICE commonly referred to as industrial emissions), the correction of the emission balance equation is not justified.
Such correction of the equation may be justified if {\it uncontrolled} emissions (in DICE commonly referred to as land emissions) need to be corrected.

According to the meaning of the equation~\EQNREF{bkm:eq1} highlighted earlier in the text, adding $x$ dollars to the consumption equation would compensate one ton of CO2 added to the emissions equation so that social utility would be kept constant. 
This newly added consumption $x$ is not caused by a change in any of the two DICE's control variables---abatement and savings rate\footnote{Savings rate is a direct equivalent of capital investment in DICE as mentioned earlier.}---and therefore is beyond DICE's control i.e. out of reach for any climate policy possibly modeled by DICE.
Moreover, since, according to DICE's concept, consumption is entirely based on economic production and regulated by DICE's decision variables---abatement and savings rate---such consumption added to the consumption equation is not supported by economy and therefore is not justified within the DICE's IAM concept.

\section{Discussion and implications for policy context}

The SCC equates additional emissions and additional consumption in a perturbed problem in such a way that the maximum societal utility in this problem remains the same as in the original (unperturbed) problem.
This SCC estimate, however, deals with uncontrolled emissions, therefore has nothing to do with any deviation of actual emissions under climate policy control from the estimated optimal plan.
In DICE's context, uncontrolled emissions are always more costly than those human-made, because, while creating economic damage via temperature increase, they are not creating any production i.e. additional consumption possibility. 
As it regards human-made emissions (or emissions under control in DICE terminology), both over-emitting (e.g. producing more economic output and/or weaker abatement) and under-emitting (e.g. producing less economic output and/or excessive abatement) as compared to the optimal level would lead to losses in utility and by that create net social cost.

The SCC calculation method implicitly relies on assumed possibility of additional consumption, which is beyond economic representation in the DICE model, and therefore no economic conclusions whatsoever can be derived from DICE employing such estimated SCC.  
From this perspective, SCC as calculated in~\EQNREF{bkm:eq1}, appears to be an irrelevant concept to justify or enforce keeping emissions at an optimal level by climate-economic policies in whatever form including SCC application as a carbon tax.

SCC only comes handy in case if, because of the reasons beyond the controls embedded in the model e.g. unforeseen disaster, the emission equation gets disturbed.
In this case, SCC can only estimate the monetary damage of such disaster in the sense that if there were an ``external'' source for increasing consumption by that amount, then that event would not create any impact on the utility.
In no case can SCC provide guidance on how to re-distribute consumption and investment after such disaster -- to answer these questions one has to carry out an optimization of the new (perturbed) problem.

Apparently, SCC and SMAC are not comparable despite they are expressed in the same units.
As the DICE model is run and the optimal solution is found, SMAC is the only optimal cost of carbon in the societal context as reflected by the models' utility function.
This social optimality is unconditional on SCC value.
SMAC ``guarantees'' the desired optimal abatement level\footnote{We would like to highlight here the issue of a surplus creation if the SMAC  value, being a marginal cost, is applied uniformly by a policy for a non-marginal abated amount. Such approach would be inconsistent with DICE's total cost of abatement valuation.}, which is conditional on technological feasibility as represented in DICE by the marginal cost of abatement specific to a time period and abatement level.

While there are cases where the numerical value of SCC in DICE happens to be close to SMAC for a relatively long time of 50-100 years after the beginning of the modeling period (see Figure~\ref{fig1} panels (a) and (b)), for other highly policy-relevant model setups with a direct temperature constraint, SMAC can be overestimated by SCC by the factor of four already in 50 years~\footnote{The 50-year period is used for SCC estimates by US EPA~\cite{refEPA2010}.} after the beginning of the modeling period (see Figure~\ref{fig1} panels (c) and (d)).
This overestimation is not conditional on any model parameters' uncertainty and stems only from the used calculation method.

To better clarify the presented SCC interpretation that refers to a perturbed problem, the ``traditional'' use of marginal values in economics can be compared to their use for the SCC calculation in DICE.
The ``traditional'' use e.g. in the producer's profit maximization problem constrained by availability of a resource required for production allows calculating the marginal price of a resource -- the maximum price that the producer would be willing to pay for one additional unit of the resource if it becomes available. 
This effectively means that the producer is an ``open system'' i.e. part of a bigger system where more resources can be made accessible e.g. by building additional mines or contracting another supplier (more detailed discussion is presented in~\cite{refSimonBlume1994} p.~452). 
Similar consideration of an ``open system'' is conceptually valid also in the climate policy context~\cite{refUzawa} where the world economy is represented by individual countries. When a particular country is considered, it can ``borrow'' from the rest of the world.
As opposed to those examples, there is just one entire world being modeled in DICE, which is the Earth's ``closed system''. This system cannot ``borrow'' from outside in terms of human-controllable emissions and economic consumption.

The findings obtained from the analysis of the SCC calculation method in the DICE model are relevant beyond the scope of DICE itself.
The discovered semantic issue is rooted in the attempt to apply the SCC value (obtained via perturbed problem describing uncontrolled emissions together with consumption not supported by economy) to shape human-controllable emissions through economic policies.
The FUND~\cite{refFUND2013,refFUNDGitHub} and PAGE~\cite{refPAGE2008,refPAGE09,refMimiPAGE,refPAGEGitHub} models that also estimate SCC, while being structurally different from DICE, both employ the same idea of an ``emission pulse'' that simply increases total emissions by adding to the emission balance equation a pre-defined amount over certain period of time to generate a new so called ``marginal'' model (that is otherwise equal to the original), which then provides a trajectory to derive the SCC value.
These newly added emissions are not caused by any change in the abatement and are the basis for the SCC estimation.
The DICE model vividly demonstrates inconsistencies resulting from the application of such constructed SCC to estimate socio-economic value of emissions under a climate policy control.

The presented analysis calls for a clear specification of the meaning of SCC in applications and suggests using the direct model output---socially optimal marginal abatement cost of carbon, SMAC---for the purposes of controlled emissions valuation while keeping in mind its validity for only incremental quantities at the optimal abatement level. SMAC with it's direct meaning has a good potential in replacing SCC to make climate policy estimates more transparent and, by doing so, facilitate humankind's progress in addressing challenges associated with global climate change.

\section*{Acknowledgments} 

The authors acknowledge early discussions around the DICE model with their IIASA colleagues Elena Rovenskaya, Artem Baklanov, Fabian Wagner, Thomas Gasser, Petr Havlik, Armon Rezai, Michael Kuhn, Stefan Wrzaczek, and others.
These discussions spurred the interest in a deeper exploration of DICE that ultimately resulted in the presented analysis.
The authors are grateful to IIASA Deputy Director General for Science Leena Srivastava for her attention and feedback.
The authors acknowledge Dr. Linus Mattauch from the University of Oxford for sharing his view on an early draft of this manuscript.
 The authors acknowledge William Nordhaus for making the DICE source code and documentation openly available as well as the clean and transparent model structure that all made the present research possible. 

\section*{Funding} 

Austrian Science Fund (FWF): P31796-N29/``Medium Complexity Earth System Risk Management'' (ERM).
European Research Council Synergy Grant number 610028 Imbalance-P: Effects of phosphorus limitations on Life, Earth system and Society (Seventh Framework Programme of the European Union).  

\section*{Author contributions} 

NK has conceptualized the problem; NK and AS have carried out the investigation; MO has contributed to funding acquisition; NK, AS, and MO have discussed the results in the process of investigation; NK has drafted the paper; all co-authors have contributed to writing the manuscript.

\section*{Competing interests} 

Authors declare no competing interests. 

\section*{Data and materials availability} 

The DICE, FUND, and PAGE models are freely available online. 

\section*{Supplementary Materials}

None.

%------------------------------------------------------------------
\setcounter{equation}{0} % numbering in appendices are independent
\newpage
\section*{Appendix A: Simplified DICE formulation}

In this section we provide the full mathematical formulation of the DICE optimization problem essentially mirroring the official DICE User's Manual~\cite{refDICEWebPage}. 
However, as it regards economic interpretation, we only present it in a very condensed simplified format as we provide only short comments on the meaning of the key variables and entirely omit the explanation of the meaning of (and the inter-relation between) the model's parameters i.e. scalar or vector constants -- all uniformly denoted by the symbols $\pi$ with the sequential numbering indicated by the lower index $i$ so that $\pi_i$ would denote a scalar constant and $\pi_i(t)$ would denote a vector constant containing individual values for each considered time period $t$. 
For brevity, we condense the time dependent arguments of a function to just the index of a time period $t$ as, for instance, in~\EQNREF{eqArgReduced} below, where the value of the function $U$ of two arguments $U[c(t),L(t)]$ (as in the DICE User's Manual~\cite{refDICEWebPage}) is simply referred to as $U(t)$.  
We keep the equations numbering compatible with the DICE User's Manual~\cite{refDICEWebPage}:

\setcounter{equation}{-1} % first equation has to have number zero
\begin{eqnarray}
	\mathrm{maximize} \quad W & & \\
	\mbox{subject to} & & \nonumber \\
	W &=& \sum_{t=1}^{T_{max}} U(t) R(t), \label{eqArgReduced} \\
	U(t) &=& \pi_1(t) \frac{ c(t)^{\pi_2} }{\pi_2}, \\
	R(t) &=& \pi_3^{-t}, \\
	Q(t) &=& [1-\Lambda(t)] \pi_4(t) K(t)^{\pi_5} \pi_6(t)^{\pi_7} /
		[1+\Omega(t)], \\
	\Omega(t) &=& \pi_8 T_{AT}(t) + \pi_9 [T_{AT}(t)]^2, \\
	\Lambda(t) &=& \pi_{10}(t) \mu(t)^{\pi_{11}}, \\
	Q(t) &=& C(t) + I(t), \label{eqConsInv} \\
	c(t) &=& C(t) / \pi_{12}(t), \label{eqExample1} \\
	K(t) &=& I(t) - \pi_{13} K(t-1), \\
	E_{Ind} (t) &=& \pi_{14}(t) [1 - \mu(t)] \pi_{15}(t) K(t)^{\pi_{16}} \pi_{17}(t)^{\pi_{18}}, \\
	\pi_{19} &\ge& \sum_{t=1}^{T_{max}} E_{Ind}(t), \label{eqIneqExample1}  \\
	E(t) &=& E_{Ind}(t) + \pi_{20}(t), \\
	M_{AT}(t) &=& E(t) + \pi_{21} M_{AT} (t-1) + \pi_{22} M_{UP} (t-1), \\
	M_{UP}(t) &=& \pi_{23} M_{AT} (t-1) + \pi_{24} M_{UP} (t-1) + \nonumber \\
		  & & {} + \pi_{25} M_{LO} (t-1), \\
	M_{LO}(t) &=& \pi_{26} M_{UP} (t-1) + \pi_{27} M_{LO} (t-1), \\
	F(t) &=& \pi_{28} \log_2 \left[ \frac{M_{AT}(t)}{\pi_{29}} \right] + \pi_{30}(t), \\
	T_{AT}(t) &=& T_{AT} (t-1) + \pi_{31} \{ F(t) - \pi_{32} T_{AT} (t-1) - \nonumber \\ 
		  & & {} - \pi_{33} [ T_{AT} (t-1) - T_{LO} (t-1) ] \}, \\
	T_{LO}(t) &=& T_{LO} (t-1) + \pi_{34} \{ T_{AT} (t-1) - T_{LO} (t-1) \}, \label{eqDiceLastEq}
\end{eqnarray}
where $W$---the objective function---is the social utility summed with discount over all time periods $t = 1,2,...,T_{max}$, and in each time period $t$ the only variables are: $U$ -- utility in the time period, $R$ -- discount\footnote{Discount is a trivial variable -- it is a time-dependent constant. We have chosen to keep it in the list with other variables only for the reasons of compatibility of equations numbering with the DICE User Manual. Otherwise, the number of variables in DICE would be reduced by one variable $R(t)$. The choice we made does not have any impact on considerations that follow.}, $Q$ -- global economic output net of damages and abatement, $\Omega$ -- represents economic damages of climate change, $\Lambda$ -- represents abatement cost, $\mu$ -- abatement level, $C$ -- total consumption, $I$ -- gross investment, $c$ -- consumption per capita, $K$ -- capital stock, $E_{Ind}$ -- industrial emissions (emissions under control),  $E$ -- total emissions, $M_{AT}$, $M_{UP}$, and $M_{LO}$ -- atmospheric, upper ocean layer, and lower ocean layer carbon respectively, $F$ -- radiative forcing, $T_{AT}$ and $T_{LO}$ -- surface temperature and the temperature of deep oceans. The initial values $K(0)$, $M_{AT}(0)$, $M_{UP}(0)$, $M_{LO}(0)$, $T_{AT}(0)$, and $T_{LO}(0)$ are known constants.

% $\pi_{19}$ --  limitation on total resources of carbon fuels (constant),
% $\pi_{20} (t)$ -- land emissions (uncontrolled emissions in DICE context, constant),

The split of the global economic product between consumption and investment~\EQNREF{eqConsInv} can be described by the ``savings rate as fraction of gross world product'' variable $S(t)$ as it is referred to in the GAMS source code of the DICE model. $S(t)$ is such that $0 \le S(t) \le 1$ and $I(t) = S(t) Q(t)$ and $C(t) = [1 - S(t)] Q(t)$, which is equivalent to~\EQNREF{eqConsInv} and therefore can be used as its replacement if the variable $S(t)$ is considered. 

The DICE optimization problem is to find such variables 
\begin{equation}\label{eqSmu}
	0 \le S(t) \le 1 \quad \mbox{and} \quad 0 \le \mu(t) \le \pi_{35}(t),
\end{equation}
which are called control variables, and also the values of other variables ($U$, $R$, $Q$, etc.) that deliver maximum of $W$ under constraints~\EQNREF{eqArgReduced} -- \EQNREF{eqDiceLastEq}. 
Equivalently, one could refrain from highlighting the $S(t)$ and $\mu(t)$ variables among others and search for the whole set of variables that deliver maximum of $W$ under constraints~\EQNREF{eqArgReduced} -- \EQNREF{eqDiceLastEq} and \EQNREF{eqSmu}.

The aforementioned constraints \EQNREF{eqSmu} are included in the GAMS DICE source code along with the constant initial values $K(0)$, $M_{AT}(0)$, $M_{UP}(0)$, $M_{LO}(0)$, $T_{AT}(0)$, and $T_{LO}(0)$ -- all appending to (Eq.~0) -- \EQNREF{eqDiceLastEq} to make the DICE optimization problem complete. 
However, for the sake of brevity and to keep compatibility with the DICE User's Manual~\cite{refDICEWebPage}, we will continue to refer to (Eq.~0) -- \EQNREF{eqDiceLastEq} as the DICE problem -- it does not affect the problem reformulation concept presented next.

\newpage
\section*{Appendix B: DICE and the standard constrained optimization problem}

Following the same approach as exercised in Appendix A to rename the DICE model's constants and vectors of constants, we can rename all model variables $U(t)$, $R(t)$, $Q(t)$, etc. denoting them using the symbol $x$ with an index, where the index is a sequential number (in any order), so that the set of the DICE's variables will become a vector $x\:=\:(x_1,x_2,...,x_n)$ of length $n$ (for brevity: $x \in \RX^n$), where $n=T_{max} \times N_V$, $T_{max}$ is the number of modeled periods, and $N_V=18$ is the number of original (not yet renamed) variables in DICE's equations (2) -- (\ref{eqDiceLastEq}). Equation (1) is defining the value $W$ which is a function of other variables -- explicitly of $U(t)$ and $R(t)$, and implicitly, through constraints (2) -- (\ref{eqDiceLastEq}) also of the others. We will rename this function to $f(x)$, where $x$ is the vector of length $n$.
Each of the equations (2) -- (\ref{eqDiceLastEq}) for each $t=1,...,T_{max}$ can be rewritten in the form $h_i(x) = 0$, where $i$ is again a sequential number and $h_i$ is a suitable function e.g.: considering $t=5$ and assuming that after applying the variables renaming procedure described here the variable $c(5)$ was renamed to $x_{123}$ and the variable $C(5)$ was renamed to $x_{456}$, and that corresponding to the equation (\ref{eqExample1}) for the time period $t=5$ the index $i$ of the respective constraint function $h$ is 78, then the function $h_{78}$ can be defined as $h_{78}(x) = x_{456} / \pi_{12}(5) - x_{123}$, where $\pi_{12}(5)$ is a known DICE constant. The equation $h_{78}(x) = 0$ in the new notation is fully equivalent to the equation (\ref{eqExample1}) in the original notation for the time period $t=5$. Similarly, for inequality constraints in DICE, there exist functions $g_j(x)$, which equivalently represent inequalities in the original DICE formulation, such as e.g.~\EQNREF{eqIneqExample1}, in the form $g_j(x) \le 0$. So, in the new notation, the DICE optimization problem is unchanged and has the form:

\begin{eqnarray}
	\mathrm{maximize}_{x \in \RX^n} \quad f(x) & & \label{eqGenOpt1}\\
	\mbox{subject to} & & \nonumber \\
	h_i(x) & = & 0, \quad i = 1,2,...,m, \\
	g_j(x) & \le & 0, \quad j = 1,2,...,k. \label{eqGenOpt3}
\end{eqnarray}
This is the standard formulation of the constrained optimization problem, which includes constraints in the form of both equalities and inequalities. This formulation is using exactly the same notation as presented in~\cite{refSimonBlume1994}. 

\newpage
\section*{Appendix C: Interpretation of marginal values in DICE}

For the DICE model in the form \EQNREF{eqGenOpt1} -- \EQNREF{eqGenOpt3} there is a theorem in~\cite{refSimonBlume1994} (19.3) for mixed constraints, which is formulated below in a simplified form considering perturbations only to equations and omitting specification of regularity assumptions.

\begin{theorem}\label{thTheorem1}
Considering the maximization problem 
\begin{eqnarray}
	\mathrm{maximize}_{x \in \RX^n} \quad f(x) & & \label{eqGenOptPert1}\\
	\mbox{subject to} & & \nonumber \\
	h_i(x) & = & a_i, \quad i = 1,2,...,m, \label{eqGenOptPert2} \\
	g_j(x) & \le & 0, \quad j = 1,2,...,k, \label{eqGenOptPert3}
\end{eqnarray}
	let denote its objective value $\max_{x \in \RX^n} f(x)$ as $V(a_1,...,a_m)$, and marginal values of its equations as $\lambda_i(a_1,...,a_m)$, then under some regularity assumptions 
\begin{equation}
	\lambda_i(a_1,...,a_m) = \frac{\partial}{\partial a_i} V(a_1,...,a_m).
\end{equation}
\end{theorem}

The perturbed DICE problem considered in the main text includes just two perturbed equations -- emissions and consumption at a time moment $t$. 
Without the loss of generality, we can assume that the respective emissions equation among the equations~\EQNREF{eqGenOptPert2} is numbered as 1 (so, it is perturbed by $a_1$ tons of CO$_2$), and the respective consumption equation is numbered as 2 (so, it is perturbed by $a_2$ dollars). 
If one would impose a requirement that these two perturbations compensate each other i.e. the original objective value does not change, that would imply

\begin{eqnarray}
	0 &=& V(a_1,a_2,0,...,0) - V(0,0,0,...,0)  
	\\ &=& (a) = \frac{\partial}{\partial a_1} V(0,0,0,...,0) \cdot a_1 
	 + \frac{\partial}{\partial a_2} V(0,0,0,...,0) \cdot a_2 \nonumber  
	\\ & & \phantom{(a) = } + \bar o(\sqrt{a_1^2 + a_2^2}) 
	\\ &=& (b) = \lambda_1(0,0,..0) \cdot a_1 + \lambda_2(0,0,..0) \cdot a_2 + \bar o(\sqrt{a_1^2 + a_2^2}).
\end{eqnarray}

Here we have employed (a) the Taylor formula, where the $\bar o(x)$ is vanishing faster than $x$ if $x$ is close to zero, and (b) the~\THEREF{thTheorem1} above. Hence, with any required accuracy, for sufficiently small $a_1$ and $a_2$ the following equation holds:

$$
\lambda_1(0,0,..0) \cdot a_1 + \lambda_2(0,0,..0) \cdot a_2 = 0.
$$

Setting $a_1 = 1$, $a_2 = x$ and assuming that these values satisfy the requirement of being "sufficiently small" and using GAMS DICE compatible notation for marginal values in the original problem $\lambda_1(0,0,..0)=eeq.m(t)$ and $\lambda_2(0,0,..0) = cc.m(t)$, the last equation becomes

$$
eeq.m(t) + x \cdot cc.m(t) = 0,
$$
which is the equation employed in GAMS DICE for SCC calculation. As it is demonstrated, this equation is stemming from perturbing the original problem by adding certain quantities to the right-hand sides of its emission and consumption equations.

\newpage
\section*{Figures}

\end{document}